\documentstyle[aps,preprint]{revtex}

\begin{document}
\begin{center}
{\Large{\bf Isospin-rich nuclei in neutron star matter}}\\

\vskip 1.0cm
Tapas Sil$^1$, J. N. De$^2$, S. K. Samaddar$^1$, X. Vi\~nas$^3$,
M. Centelles$^3$,\\ B. K. Agrawal$^{1,4}$ and S. K. Patra$^5$ \\
$^1$Saha Institute of Nuclear Physics, 1/AF Bidhannagar, Kolkata 700064, India\\
$^2$Variable Energy Cyclotron Centre, 1/AF Bidhannagar, Kolkata 700064, India\\
$^3$Departament ECM, Facultat de F{\'\i}sica, Universitat de
Barcelona, Diagonal 647, 08028 Barcelona, Spain\\
$^4$The Cyclotron Institute, Texas A$\&$M University, College Station,Texas
77843,USA\\
$^5$Institute of Physics Sachivalaya Marg, Bhubaneswar, 751005, India
\end{center}

\begin{abstract}
Stability of nuclei beyond the drip lines in the presence of an
enveloping gas of nucleons and electrons, as prevailing in the inner
crust of a neutron star, is studied in the temperature-dependent
Thomas-Fermi framework. A limiting asymmetry in the isospin space
beyond which nuclei cannot exist emerges from the calculations. The
ambient conditions like temperature, baryon density and neutrino
concentration under which these exotic nuclear systems can 
be formed are studied in some detail.
\end{abstract}

\vskip 1.0cm
PACS Number(s): 21.10.-k,21.60.-n,97.60.Jd,26.60.+c
\newpage
\section{Introduction}

In the outer region of the core of a neutron star, at densities $\sim
1.5 \times 10^{14}$ g/cm$^3$ or more, nucleons are distributed
uniformly forming a homogeneous system \cite{bay,buc,pet,hei}. At
lower densities, in the inner crust of the star, inhomogeneities
appear and then the positive charges get concentrated in individual
nuclei of charge $Z$ which are embedded in a sea of neutrons,
electrons and possibly neutrinos, the whole system being charge
neutral. The nuclei arrange themselves in a periodic
lattice \cite{sha} due to the electrostatic interaction. In the
typical conditions prevailing in the neutron star interior, the nuclei
are in complete thermodynamic equilibrium with the environment and are
assumed to be in $\beta$-equilibrium. They may well be beyond the
neutron drip line \cite{bay,lan} known for laboratory nuclei with
$(N-Z)/A \leq 0.35$. The excess neutrons of the very neutron rich
nuclei present in the inner neutron star crust, which would otherwise
decay under terrestrial conditions, are held together in stable
equilibrium by the pressure exerted by the surrounding neutron sea.

The properties of the isospin-rich nuclei may be quite different from
those of the terrestrial nuclei. For example, with increasing density,
the nuclei may pass through different exotic shapes, namely from
spheres to cylinders, slabs, cylindrical holes and spherical bubbles
\cite{rav,oya}. The presence of nonspherical nuclei could affect
significantly the pinning of vortices and neutrino emission from
neutron stars \cite{lor}. The existence of these exotic-shaped nuclei
is, however, model-dependent \cite{pet,che}. The external gas
surrounding the nuclei may also influence the density distribution of
these nuclei and thus may modify their size \cite{bay}.

The nuclear equation of state (EOS) plays the pivotal role in
determining the macroscopic properties of a neutron star, such as its
mass, radius and moment of inertia \cite{eng}. In this context, the
presence of different nuclear species in the outer and inner crustal
regions provides the starting point for the consideration of some
important aspects, such as its superfluid and elastic properties. On the
other hand, for nuclei immersed in a nucleonic gas there are two basic
concerns from a microscopic viewpoint. One, a thermodynamically
consistent treatment of the coexistence of the two phases of nuclear
matter (namely, the nuclear liquid and the surrounding gas)
\cite{lat1,lam,mul}, the other, a plausible description of the
interface between the liquid and the gas \cite{lat,mye,cen}. It has
been shown that this problem can be taken care of by solving the
density profile in the subtraction procedure of Bonche, Levit and
Vautherin (BLV) \cite{bon,sur}. In this method, the density profile of
the liquid-plus-gas system and that of the gas are solved in a
self-consistent procedure and then the extensive observables referring
to the liquid are obtained as the difference between those of the two
solutions. The influence of the surrounding gas on the surface energy
is then automatically taken into account. 
The same concept can be applied to situations at zero temperature
where drip nucleons occur and the system coexists with an outer
nucleonic phase, as shown for very asymmetric cold semi-infinite
nuclear matter in Ref.\ \cite{cen}.

In this paper we focus on the effect of the external gas on the
structure and stability of finite nuclei with large neutron excess, to
have a broader understanding of the conditions under which these
nuclei may exist in neutron star matter. A preliminary study for
isolated nuclei immersed in a nucleonic gas at zero temperature has
been done recently \cite{jde}; in the present work we extend these
ideas in the appropriate astrophysical context. We also extend the
calculations to non-zero temperature which is relevant at the
formation stage of the neutron stars.

We assume that in the nuclear matter at sub-nuclear density the nuclei
are located in a lattice. In order to simplify our calculation, the
Wigner-Seitz approximation \cite{sha} 
is applied, where each lattice volume is
replaced by a spherical cell of radius $R_c$, the nucleus being
located at its centre. The matter in each cell is taken to be charge
neutral, ie, the number of electrons is equal to the number of protons
in the cell. The neutrino density in this cell is determined from the
$\beta$-equilibrium condition. For a given average density of nuclear
matter with a certain proton concentration, the density distribution
is solved self-consistently. The calculations are performed in the
finite temperature Thomas-Fermi formalism. The matter in the cell in
which the protons and neutrons coexist does not define the nucleus
itself, the nucleus is identified after subtraction of the gas part
generated self-consistently, as in previous investigations of
excited nuclei or asymmetric semi-infinite matter \cite{cen,bon,sur}.

In Section II, the model employed in the calculations is introduceed.
The results and discussions are presented in Section III\@. Section IV
contains the concluding remarks.

\section{Model}

In the inner region of the crust of a neutron star, we consider a mixture of
neutrons, protons (some of which may exist as bound nuclear clusters),
electrons and neutrinos in thermodynamic equilibrium. We contemplate
both cold matter as well as matter at a finite temperature. We ignore
the contributions from alpha particles and also from photons. We
further ignore the plasma effects. All these effects are known to be
rather small \cite{mpi}. The nuclear clusters are assumed to be
arranged in a body-centered cubic lattice which we approximate by
Wigner-Seitz (WS) cells defined as spheres with radius $R_c$. Each
cell is assumed to be electrically neutral and interactions among the
cells are neglected. At the densities that we are interested in, the
Fermi momenta of the electrons are much larger than their rest
mass, the electrons are then extremely relativistic and can be assumed
to be uniformly distributed in the cell. We assume the matter to be in
$\beta$-equilibrium, ie, the chemical potentials of the considered
species fulfil
\begin{equation}
\label{beta}
\mu_n=\mu_p+\mu_e-\mu_\nu .
\end{equation}

At a fixed temperature, under the condition of 
charge neutrality, the variables in the problem
are the average baryon density $\langle\rho\rangle$, the proton
fraction $Y_p= Y_e$ (the electron fraction), the neutrino fraction
$Y_{\nu}$ and the lattice radius $R_c$. With the $\beta$-equilibrium
condition given by Eq.\ (\ref{beta}), if $\mu_n$, $\mu_p$ and $\mu_e$
are known, then $\mu_\nu$ can be determined which in turn yields
$\rho_\nu$ when the neutrinos are taken to be a degenerate Fermi gas.
Then only three of the variables are left independent. The condition
of $\beta$-equilibrium of the baryons with the electrons and neutrinos
guarantees that the system has the minimum free energy. If one imposes
further the constraint that the lattice sites contain a particular
nuclear species with a given charge $Z_{cl}$ and baryon number
$A_{cl}$ (immersed in a nucleonic gas or not), then there is only one
independent variable.

To obtain the thermodynamic properties of the system of baryons,
electrons and neutrinos, we minimize the free energy of the system in
the WS cell with the constraint of number conservation of the
individual species. Under the conditions of charge neutrality and
$\beta$-equilibrium, the relevant grand potential is given by
\begin{equation}
\label{gpdef}
\Omega = {\cal F}-\sum_q\mu_qA_q,
\end{equation}
where ${\cal F}$ is the free energy and the index $q=(n,p)$ refers to
neutrons and protons. The free energy has the following expression: 
\begin{equation}
\label{fdef}
{\cal F}(\langle\rho\rangle,Y_p,Y_{\nu},T)=\int \left[{\cal H}(r)-Ts(r)
+ {\cal E}_c(r) +f_e(\rho_e)+f_{\nu}(\rho_\nu)\right] d{\bf r} .
\end{equation}
In Eq.\ (\ref{fdef}) the integration is over the volume of the WS
cell. Here ${\cal H}$ refers to the baryonic energy density
excluding the Coulomb energy, $s$ is the entropy density of the
baryons, ${\cal E}_c$ the Coulomb energy density of the system and
$f_\nu$, $f_e$ are the free energy densities of the neutrinos and of
the electrons (Coulomb interaction excluded). 

For the nuclear energy density, we use the Skyrme energy density
functional. It is written as
\begin{eqnarray}
\label{skme}
{\cal H}(r)&=&\frac{\hbar^2}{2m_n^*}\tau_n 
+\frac{\hbar^2}{2m_p^*}\tau_p+
\frac{1}{2}t_0\left[\left(1+\frac{x_0}{2}\right)\rho^2-\left(x_0+
\frac{1}{2}\right)\left(\rho^2_n+\rho^2_p\right)\right]\nonumber\\
&&-\frac{1}{16}\left[t_2\left(1+\frac{x_2}{2}\right)
-3t_1\left(1+\frac{x_1}{2}\right)\right](\nabla \rho)^2\nonumber\\
&&-\frac{1}{16}\left[3t_1\left(x_1+\frac{1}{2}\right)
+t_2\left(x_2+\frac{1}{2}
\right)\right]\left[(\nabla\rho_n)^2+(\nabla\rho_p)^2\right]\nonumber\\
&&+\frac{1}{12}t_3\rho^\alpha\left[\left(1+\frac{x_3}{2}\right)\rho^2
-\left(x_3+\frac{1}{2}\right)\left(\rho_n^2+\rho_p^2\right)\right],
\end{eqnarray}
where $\rho=\rho_n+\rho_p$ and the effective mass of the nucleons 
$m_q^*$ is defined through
\begin{eqnarray}
\label{effm}
\frac{m}{m_q^*(r)}&=&1+\frac{m}{2\hbar^2}\left\{\left[t_1\left(1+
\frac{x_1}{2}\right)
+t_2\left(1+\frac{x_2}{2}\right)\right]\right.\rho \nonumber\\
&& +\left.\left[t_2\left(x_2+\frac{1}{2}\right)
-t_1\left(x_1+\frac{1}{2}
\right)\right] \rho_q\right\}.
\end{eqnarray}
In the numerical calculations we shall employ the SKM* interaction,
whose parameters can be found in Ref.\ \cite{bra}. In the Thomas-Fermi
approximation the kinetic energy density $\tau_q$ is given by
\begin{eqnarray}
\label{ke}
\tau_q(r) &=&\frac{3}{5}(3\pi^2)^{2/3}\rho_q^{5/3}
\qquad\qquad\qquad \ \
\mbox{at } T=0 ,
\\
\tau_q(r) &=& \frac{1}{2\pi^2}
\left( \frac{2 m_q^* T}{\hbar^2} \right)^{5/2} J_{3/2}(\eta_q)
\qquad
\mbox{at } T\ne0 .
\label{ket}
\end{eqnarray}
%
The fugacity $\eta_q$ is obtained as
\begin{equation}
\label{fug}
\eta_q(r)=(\mu_q-V_q(r))/T,
\end{equation}
where $V_q$ is the single-particle potential experienced by nucleons
(including the Coulomb part for the protons). The nucleonic density
$\rho_q$ is related to $\eta_q$ by
\begin{equation}
\label{den}
\rho_q(r)=  \frac{1}{2\pi^2}
\left( \frac{2 m_q^* T}{\hbar^2} \right)^{3/2} J_{1/2}(\eta_q).
\end{equation}
The functions $J_k(\eta_q)$ in Eqs.\ (\ref{ket}) and (\ref{den}) are
the standard Fermi integrals. 

For the entropy density of the nucleons one has
\begin{equation}
\label{ent}
s(r)= \sum_q [(5/3)J_{3/2}(\eta_q)/J_{1/2}(\eta_q) - \eta_q] 
\, \rho_q .
\end{equation}
The direct part ${\cal E}_c^d$ of the Coulomb energy density 
${\cal E}_c$ of the charged particles is
\begin{equation}
\label{cde}
{\cal E}_c^d(r)=\frac{1}{2} (\rho_p(r)-\rho_e) \int
(\rho_p(r^{\prime})-\rho_e)\frac{e^2}{\mid{\bf r}-
{\bf r^{\prime}}\mid} d{\bf r^{\prime}} ,
\end{equation}
while the exchange part is computed in the Slater approximation:
\begin{equation}
\label{cee}
{\cal E}_c^{ex}(r)=-\frac{3}{4}\left(\frac{3}{\pi}\right)^{1/3}e^2
(\rho_p^{4/3}(r)+\rho_e^{4/3}).
\end{equation}
For $f_e$ and $f_\nu$ we use the standard expressions
\cite{cox}. For the minimisation procedure, we take recourse to the
finite temperature Thomas-Fermi approximation which yields Eq.\
(\ref{fug}). This equation is solved self-consistently leading to the
density distributions of neutrons and protons in the spherical WS
cell. 

Under the conditions of interest here, the baryonic fluid in the WS
cell can congregate into nuclei (we consider only the spherical
shapes) located at the centre of the cell and may be embedded in a low
density gas of nucleons. Near the edge of the cell, for hot or for
isospin rich systems beyond the drip line, the baryon density profile
is found to be practically uniform which is identified as the low
density gas. To delineate the nucleus from the embedding low-density
environment, we follow the BLV procedure which has been used earlier
in the Hartee-Fock (HF) framework \cite{bon} as well as in a
Thomas-Fermi (TF) scheme \cite{sur} for a hot nucleus in coexistence
with the vapour surrounding it. The method is based on the existence
of two solutions to the HF or TF equations, one corresponding to the
liquid phase with the surrounding gas (LG) and the other corresponding
to the gas (G) alone.

For an isolated hot nucleus in equilibrium with the gas surrounding
it, in the absence of Coulomb forces, the densities $\rho_{LG}^q$
and the gas density $\rho_G^q$ can be obtained from the variational
equations 
\begin{eqnarray}
\label{omev}
\frac{\delta\Omega_{LG}^N}{\delta\rho_{LG}^q}=0\\
\frac{\delta\Omega_{G}^N}{\delta\rho_{G}^q}=0
\end{eqnarray}
where ${\Omega_{LG}^N}$ and ${\Omega_{G}^N}$ are the 
grand potentials of the respective systems with the
same chemical potential. In our procedure, the solution for $\rho_L =
\rho_{LG}-\rho_G$, which may be called the liquid profile, is
independent of the box volume in which the calculation is done. In the
presence of the Coulomb force, however, the Coulomb repulsion
increases with the box size, driving the protons to the edge of the
box and leading ultimately to a divergence problem. In order to work
out a convergent prescription, Bonche {\it et al.} \cite{bon}
calculated the Coulomb energy $E_c$ from the subtracted proton density
$\rho_{LG}^p-\rho_G^p$ and obtained the density profile from the
variation of the subtracted grand potential ${\bar
\Omega}=\Omega_{LG}^N-\Omega_G^N+E_c$ with respect to both $\rho_{LG}^q$
and $\rho_G^q$.

In the astrophysical context, the situation is however different due
to the presence of electrons. The Coulomb energies for the LG and G
phases are
\begin{eqnarray}
\label{cdelg}
E_{LG}^c&=& \frac{1}{2} \int (\rho_{LG}^p(r^\prime)-\rho_e)
\frac{e^2}{\mid{\bf r}-{\bf r^{\prime}}\mid} (\rho_{LG}^p(r)-\rho_e)
d{\bf r} d{\bf r^\prime},\\
E_{G}^c&=& \frac{1}{2} \int (\rho_{G}^p(r^\prime)-\rho_e)
\frac{e^2}{\mid{\bf r}-{\bf r^{\prime}}\mid} (\rho_{G}^p(r)-\rho_e)
d{\bf r} d{\bf r^\prime}\nonumber\\
&&+ \int \rho_{L}^p(r^\prime) \frac{e^2}{\mid{\bf r}
-{\bf r^{\prime}}\mid} 
(\rho_{G}^p(r)-\rho_e) d{\bf r} d{\bf r^\prime}.
\label{cdeg}
\end{eqnarray}
The direct part of the single-particle Coulomb potentials ($\delta
E_{LG}^c / \delta \rho_{LG}^p$ and $\delta E_{G}^c / \delta \rho_{G}^p$)
for the LG are the G solutions is the same. It is given by
\begin{equation}
\label{cdsp2}
V_c^d(r)= \int (\rho_{LG}^p(r^\prime)-\rho_e)
\frac{e^2}{\mid{\bf r}-{\bf r^{\prime}}\mid} d{\bf r^\prime}.
\end{equation}
Since the system is charge neutral, the divergence problem does not
arise. The solutions to the density profiles can be directly obtained
from the variations of the total (with Coulomb) grand potentials
$\Omega_{LG}$ for the liquid-plus-gas phase and $\Omega_G$ for the gas
phase with respect to $\rho_{LG}^q$ and $\rho_G^q$, respectively. The
resulting coupled equations are
\begin{eqnarray}
\label{tfet1}
T\eta_{LG}^q(r)+V_{LG}^q+V_{LG}^c(\rho_{LG}^p,\rho_e)&=&\mu_q,\\
T\eta_{G}^q(r)+V_{G}^q+V_{G}^c(\rho_{LG}^p,\rho_G^p,\rho_e)&=&\mu_q .
\label{tfet2}
\end{eqnarray}
At zero temperature they reduce to
\begin{eqnarray}
\label{tfe01}
(3\pi^2)^{2/3}\frac{\hbar^2}{2m_q^*}(\rho_{LG}^q)^{2/3}+V_{LG}^q+
V_{LG}^c(\rho_{LG}^p,\rho_e)&=&\mu_q,\\
(3\pi^2)^{2/3}\frac{\hbar^2}{2m_q^*}(\rho_{G}^q)^{2/3}+V_{G}^q+
V_{G}^c(\rho_{LG}^p,\rho_G^p,\rho_e)&=&\mu_q.
\label{tfe02}
\end{eqnarray}
In Eqs.\ (\ref{tfet1})--(\ref{tfe02}), $V_{LG}^q$ and $V_G^q$ refer to
the nuclear part of the single-particle potential corresponding to the
liquid-plus-gas and the gas solutions. $V_{LG}^c$ ($V_G^c$) is the sum
of the direct part of the Coulomb single-particle potential given by
Eq.\ (\ref{cdsp2}) and the exchange term $-e^2 (3/\pi)^{1/3}(\rho_{LG
(G)}^p)^{1/3}$ for the protons in the LG (G) phase. At zero
temperature, if the cluster is within the drip lines, the cell does
not contain any nucleonic gas. Equations (\ref{tfe01})--(\ref{tfe02})
then automatically lead to a gas solution which is zero throughout.

The nucleonic chemical potential is determined from the constraint of
nucleon number conservation. In the case of a given nuclear cluster
with neutron number $N_{cl}$ and proton number $Z_{cl}$ embedded in a
nucleonic gas, the conservation of the number of nucleons in the
cluster requires [from Eqs.\ (\ref{tfet1}) and (\ref{tfet2})]
\begin{eqnarray}
\label{mucal}
\mu_q&=&\frac{1}{A_q}\left\{\int \left[
T\eta_{LG}^q(r)+V_{LG}^q(r)+V_{LG}^c(r)\right]\rho_{LG}^q(r) 
d{\bf r}\right.
\nonumber\\ && \left. - \int \left[ T\eta_{G}^q(r)
+V_{G}^q(r)+V_{G}^c(r)\right] \rho_{G}^q(r) d{\bf r}\right\} .
\end{eqnarray} 
Here $A_q$ refers to $N_{cl}$ or $Z_{cl}$. An equation similar to
(\ref{mucal}) follows from Eqs.\ (\ref{tfe01}) and (\ref{tfe02}) at
zero temperature.

On the other hand, when $\langle\rho\rangle$ is given in a WS cell
with a given proton concentration, the total number of neutrons $N$
and protons $Z$ in the cell is defined. Then, the chemical potential
is obtained as
\begin{equation}
\label{muav}
\mu_q=\frac{1}{A_q}\int \left[ T\eta_{LG}^q(r)+V_{LG}^q(r)
+V_{LG}^c(r)\right] \rho_{LG}^q(r) d{\bf r} ,
\end{equation}
where now $A_q$ refers to $N$ or $Z$.

\section{Results and discussion}

In an earlier paper by some of the present authors 
\cite{jde}, calculations at
zero temperature on some structural properties of isolated nuclei much
beyond the drip line were reported. In that case, the excess pressure
exerted by the enveloping gas stabilises the nucleus even beyond the
nominal drip lines (which are defined by $\mu_n= 0$ or $\mu_p=0$ in
the Thomas-Fermi approach). We now place these calculations in a
broader context related to the environment existing in the inner crust
of a neutron star. We take cognizance of the presence of electrons and
neutrinos existing in $\beta$-equilibrium with the neutrons and
protons. We extend the calculations also at finite temperature. As
mentioned earlier, the calculations are performed in the finite
temperature Thomas-Fermi framework in a Wigner-Seitz lattice employing
the SKM* interaction. For completeness, we also discuss the situation for
the proton-rich nuclei though they may not exist in the crustal matter of
the neutron stars.

In the calculations on asymmetric infinite and semi-infinite nuclear
matter in equilibrium with a drip phase, it was observed
\cite{lat1,mye,cen} that the neutron-proton asymmetry could be
increased arbitrarily till the two phases merge into a uniform system
when the densities and the asymmetries of both phases become equal. On
the contrary, in the self-consistent TF calculation for isolated
finite nuclei, it was found that one cannot add or remove neutrons
from nuclei arbitrarily \cite{jde}.
There exists a limiting neutron-proton asymmetry
$I=(N_{cl}-Z_{cl})/A_{cl}$  beyond which the system becomes
thermodynamically unstable; establishing chemical equilibrium between
the nuclear phase and the gas phase further becomes impossible.

In the realistic conditions considered in the present investigation,
including the lattice effects, we calculate the stability limits at
$T=0$ as shown in the upper panel of Fig.\ 1.
We find that both the limiting asymmetry (full line) and
the drip asymmetry (dashed line) are not very sensitive to the atomic
number $Z_{cl}$ of the nucleus. The influence of the lattice electrons
on both the drip lines and the limiting lines for neutrons and protons
is manifested in the lower panel of Fig.\ 1. The neutron lines are
nearly unaffected. The proton lines are however influenced
significantly, particularly for the heavier clusters. Both the proton
drip and limiting lines get extended with inclusion of the lattice
effects because of the dilution of the Coulomb force in the presence
of electrons. The calculations are done in a lattice size $R_c=15$ fm.
The results are found to be rather insensitive to the size of the 
lattice.

The influence of temperature on the drip and limiting lines, taking
into account the lattice effects, is displayed in Fig.\ 2. There, we
compare the results obtained at $T=0$ and at $T=6$ MeV\@. Both the
neutron and proton drip lines are extended at finite temperature as
found earlier in calculations for isolated nuclei \cite{bes,jde1}. On
the other hand, the neutron limiting line shrinks with
temperature, whereas the proton limiting line remains essentially
unaffected. 

In Fig.\ 3 we present the density profile of neutrons (left panel) and
protons (right panel) for an extremely neutron-rich nucleus
$^{330}$Pb 
(the neutron drip line is located at $^{276}$Pb).
We display calculations for both the case with lattice and
the case without lattice (isolated nuclei) at $T=0$. At a finite
temperature $T=6$ MeV we show the calculations only with lattice
effects. 
The total density profile $\rho_{LG}^q$ of the liquid-plus-gas phase
is shown in the upper panels. For neutron-rich nuclei the influence
of the electrons in the lattice is found to have only a nominal effect
on the density distributions. The finite neutron density at the cell
boundary even at $T=0$ reflects the presence of the neutron gas for a
nucleus beyond the neutron drip line. With increasing temperature, the
central density is depleted with the appearance of a thickening tail.
The gas densities $\rho_G^q$ are shown in the bottom panels and the
liquid densities $\rho_L^q$ obtained as the difference between
$\rho_{LG}^q$ and $\rho_G^q$ are shown in the central panels.

The density profile $\rho_L^q$ of the isolated cluster $^{330}$Pb is
found to be independent of the size of the box. However, in a lattice,
because of the modification of the Coulomb force in the presence of
the electrons, the nucleonic densities in the cluster are box-size
dependent, though weakly. The liquid density vanishes at a distance $r
< R_c$ even at $T=6$ MeV\@. The neutron gas density is found to be
practically constant throughout the box; at finite temperature,
the gas density is larger as expected. For this neutron-rich system,
at $T=0$, the proton gas does not exist; however, at finite
temperature ($T=6$ MeV), a few protons are present in the gaseous
state. In finite temperature calculations for isolated nuclei, the
proton gas density profile is strongly polarised \cite{bon} due to the
repulsion from the nuclear core. In the presence of the lattice
electrons, the density polarisation exists only in the vicinity of the
core, at further distances the proton gas density is found to attain a
nearly constant value.

The density distributions for an extremely proton-rich isotope
$^{140}$Pb are shown in Fig.\ 4. The proton and neutron distributions
at $T=0$ (both with and without lattice contributions) and at $T=6$
MeV with lattice effects are displayed in the left and right panels,
respectively. The electrons in the lattice dilute the Coulomb
repulsion thereby lowering the proton chemical potential, particularly
for proton-rich heavy nuclei, as an effect of which the proton drip
line is extended. For instance, for an isolated
Pb nucleus the proton drip line is located at $^{182}$Pb; the
influence of the lattice electrons pushes it to $^{143}$Pb. 
The total (LG) proton density distribution for the isolated $^{140}$Pb
nucleus at zero temperature indicates the presence of a proton gaseous
phase. This proton gas is strongly polarised as is clear from the
proton gas density distribution shown in the bottom panel. With
inclusion of the lattice electrons, since the $^{140}$Pb nucleus is
just beyond the modified drip line, the proton gas density is
extremely dilute as is evident from the bottom left panel. The central
panel of Fig.\ 4 indicates that the liquid density profiles with and
without lattice corrections at $T=0$ are not much different. The
influence of the lattice electrons on the neutron density distribution
is negligible as displayed in the right panel. For the present system,
there is no neutron gas at zero temperature. The characteristics of
the density distributions at finite temperature are found to be
basically the same as discussed in the context of a neutron-rich
nucleus in Fig.\ 3.

All of the subsequent calculations we shall present are performed
including the influence of the lattice electrons. In Fig.\ 5, the rms
neutron and proton radii of the lead isotopes 
are displayed for $T=0$ as well as for  $T=6$ MeV\@.
At zero temperature, except at the edges of limiting asymmetry, the
rms radius changes almost linearly with the mass number. The faster
change in radius near the edges points to the onset of instability.
Similar observations were also noted in the calculations without
lattice effects \cite{jde}. Temperature increases the rms radii, its
effect being more pronounced near the stability limit. By comparing
the upper and lower panels of Fig.\ 5 (neutrons and protons), one
observes the growth of the neutron or proton skins with positive or
negative asymmetry, the effect being more noticeable at zero than at
finite temperature.

We now explore the ambient conditions in which a particular nucleus
can exist at a certain temperature in the inner crust of the neutron
star. For this purpose, we have chosen $^{80}$Ca and $^{170}$Sn as
representative systems. The nucleus $^{80}$Ca is well beyond the
Thomas-Fermi drip line ($^{68}$Ca being on the drip line) whereas
$^{170}$Sn is just at the drip boundary. As mentioned in Sec.\ II,
among the four variables $\langle\rho\rangle$, $Y_p$, $Y_\nu$ and
$R_c$, only one is independent once the $N_{cl}$ and $Z_{cl}$ values
of the nucleus are constrained and $\beta$-equilibrium assumed. For
the existence of the particular isotope the values of the rest of the
variables are then fixed. The correlations among $\langle\rho\rangle$,
$Y_p$ and $R_c$ are displayed in Fig.\ 6 for the aforementioned two
isotopes at $T=0$, 3 and 6 MeV\@. The filled circles correspond to
different values of $R_c$ at which the calculations have been
performed. The values of $R_c$ range from 12 fm to 30 fm at a step of
3 fm, increasing with decreasing $\langle\rho\rangle$ for all the
cases shown. The nucleus $^{80}$Ca is beyond the neutron drip line, it
is hence embedded in a neutron sea even at zero temperature.
For the isolated nucleus, the density of the neutron gas is independent of
the box size. Even in
calculations with inclusion of effects from lattice electrons, the
neutron chemical potential $\mu_n$ and hence $\rho_G^n$ change very
little with the cell size. 
 This explains the gradual fall of
$\langle\rho\rangle$ and $Y_p$ with increasing $R_c$. For $^{170}$Sn,
since the nucleus is at the neutron drip boundary, the fall of
$\langle\rho\rangle$ with increasing $R_c$ is vertical at $T=0$ as
there is no neutron gas in the cell. At finite temperature, however, the
cluster becomes embedded in a nucleonic gas of mostly neutrons and some
protons causing a reduction in $\langle\rho\rangle$ and $Y_p$ with
increasing $R_c$.

The ambient conditions for stability are investigated further for Pb
isotopes ranging from $A_{cl}=110$ to $A_{cl}=330$ at the same three
temperatures mentioned for $^{80}$Ca and $^{170}$Sn. The calculations
for Pb have been performed at a the value of cell size $R_c=15$ fm.
The results are displayed in Fig.\ 7. The proton drip point for lead
at $T=0$ is at $A_{cl}=143$, while the neutron drip point is at
$A_{cl}=276$. These points are indicated by the vertical arrows in the
figure. In the upper panel of this figure, the average density
$\langle\rho\rangle$ is plotted against mass number $A_{cl}$. The
curve for the zero temperature case has a minimum at $A_{cl}\sim 143$
which is the proton drip point. Between the drip points ($A_{cl}=143$
to 276), since there is no nucleonic gas around at $T=0$, the average
density $\langle\rho\rangle$ within the WS cell increases linearly
with mass number. Beyond the drip points, the nuclei are surrounded by
a nucleonic gas for their stability, $\langle\rho\rangle$ therefore
increases faster. For a hot nucleus there are evaporated nucleons in
the cell and the average density is larger.

The proton fraction $Y_p$ as a function of the mass number is shown in
the middle panel of Fig.\ 7. A kink is seen to occur at $A_{cl}=143$, the
proton drip point below which $Y_p$ rises faster because of the
presence of the proton gas. For $A_{cl}>143$, the charge $Z_{cl}$ is fixed.
Hence, with increase of mass number, $Y_p$ decreases linearly up to
the neutron drip point beyond which the slope of the curve changes as
neutrons appear in the drip phase. With temperature, because of
evaporation of nucleons taking place, the kinks smoothen out, but the
overall qualitative features are the same. The lower values of $Y_p$
at finite temperature reflect the predominance of neutrons in the gas
for isotopes that are not very proton-rich. The variation of the
neutrino fraction $Y_\nu$ ($=\rho_\nu/\langle\rho\rangle$) with the
isotope mass is displayed in the bottom panel. With increasing mass
number, it is found to decrease. The effect of temperature on $Y_\nu$
is very weak. Inspection of Fig.\ 7 reveals that for a particular
$\langle\rho\rangle$, $A_{cl}$ is generally double-valued. If the isotope
is proton-rich, $Y_p$ is necessarily large and $Y_\nu$ also comes out
to be large. For the same value of $\langle\rho\rangle$, the
neutron-rich isotope is formed when $Y_\nu$ is small.

So far our calculations have been done by fixing a given cluster
($N_{cl}$, $Z_{cl}$), which yields various sets of values for the
parameters of the problem. A common practice is, however, to fix the
parameters like $\langle\rho\rangle$ and $Y_p$ and study the change in
the internal structure of neutron star matter. When
$\beta$-equilibrium is assumed, for a given value of $\langle\rho\rangle$ and
$Y_p$, the free energy density (equivalent to free energy per baryon in the
present case) in the system of nucleons, electrons and neutrinos is a
minimum in any given cell size $R_c$. The cell size sets the
periodicity in the neutron star matter; it is not a priori known and
therefore is taken as a parameter. Even for  given values of
$\langle\rho\rangle$ and $Y_p$, the neutrino concentration $Y_\nu$
(under $\beta$-equilibrium conditions) would be different for
different values of $R_c$. One may look for a global minimum in the
free energy density by varying $Y_\nu$ (through variation of $R_c$).
This is displayed in Fig.\ 8 for $T=0$ and for three values of
densities keeping the proton concentration $Y_p$ fixed at 0.3. The
minima in the free energy density are marked by arrows. With
decreasing density, the cell size corresponding to the global minimum
increases. 

To understand the origin of the minimum in the free energy profile
presented in Fig.\ 8, the variation of its different components with
lattice size for two values of $\langle\rho\rangle=0.001$ (left panel)
and 0.01 fm$^{-3}$ (right panel) are shown in Fig.\ 9. Here $Y_p$ is
fixed at 0.3. Since these calculations are carried out at zero
temperature, the total energy per baryon $e_{tot}=F/A$ is given by
\begin{equation}
F/A=e_N+e_{lat}+e_e+e_{\nu} .
\end{equation}
The different contributions $e_N$, $e_{lat}$, $e_e$ and $e_{\nu}$ are
the nuclear energy including the Coulomb interaction among the
protons, the lattice energy (ie, the Coulomb interaction energy due to
the presence of the electrons), the electron kinetic energy and the
neutrino energy per baryon, respectively. It is seen that $e_N$ has a
minimum at a lattice radius $R_c$ somewhat smaller compared to
$R_c^{min}$, the value of $R_c$ at which $F/A$ is minimum in Fig.\ 8
at the corresponding $\langle\rho\rangle$. This may be understood from
the variation of the cluster size $A_{cl}$ with $R_c$ as shown in the
middle panels of the figure. The cluster size increases monotonically
with $R_c$ and for these neutron-rich clusters the minimum binding
energy per nucleon occurs at around $A_{cl} \sim 100$ (for nuclei on the
$\beta$-stability line, this occurs for $^{56}$Fe). The lattice energy
$e_{lat}$ decreases monotonically with $R_c$. This is because for
given $Y_p$, the proton number and hence the electron number increases
with increasing lattice size keeping the electron density constant, as
a result of which $e_{lat}\sim -A^{2/3}$ \cite{sha}. The electron
kinetic energy per baryon $e_e$ is a constant for a given density and
$Y_p$ and therefore is not shown in the figure. The neutrino energy
$e_{\nu}$ passes through a minimum at a value of $R_c$ somewhat
smaller than $R_c^{min}$ where the neutrino fraction $Y_{\nu}$ is also
a minimum (shown in the bottom panel). The competition between
$e_{lat}$ and the rising parts of ($e_N$ + $e_{\nu}$) determines the
location of the minimum in the total free energy at $T=0$.

In Fig.\ 10, we display the thermal evolution of the cluster
composition ($A_{cl},Z_{cl}$) at different fixed densities, at a
particular value of $Y_p=0.3$ and for a lattice size $R_c=15$ fm. The
number of neutrons and protons in the WS cell is then fixed. The full
lines correspond to the mass number $A_{cl}$ of the cluster and the
dotted lines refer to its charge. For the chosen conditions, the
clusters formed at zero temperature in the WS cell comprise all the
nucleons without any gas. As the temperature rises, the cluster size
shrinks and becomes surrounded by the gas of the evaporated nucleons,
the nuclear liquid and the gas being in thermodynamic equilibrium. The
cluster evaporates completely at a particular temperature depending on
the chosen density, and then the cell contains only the gas of nucleons
and  the leptons. We call it the boiling
temperature $T_B$. As the density $\langle\rho\rangle$ is decreased,
the temperature $T_B$ falls down. A similar situation was observed in
the context of the liquid-gas phase transition in finite nuclei, where
the phase transition temperature was found to decrease \cite{jde2}
with an increase in the so-called freeze-out volume or 
a lowered average density
$\langle\rho\rangle$. Figure
10 also tells us about the appearance of different nuclear clusters as
the neutron star evolves in the formation stage. Initially, the
temperature $T$ may be as high as $\sim 10$ MeV and then all the
nucleons will likely be in the gas phase. As time flows by, the system
cools down and seeds of nuclei with increasing size start appearing
from the gas.

\section{Conclusions}

We have investigated the structural properties of nuclei and nuclear
matter as can be found in the environment of the inner crust of a
neutron star. These nuclei are dipped in a sea of electrons, neutrinos
and of low-density nucleons. 
Because of the pressure exerted by the surrounding nucleonic gas,
nuclei may exist even far beyond the nominal nuclear drip line.
Following the BLV subtraction procedure, we give a prescription in a
Thomas-Fermi  framework to properly isolate the nucleus from its
environment. This is extremely helpful in further exploring the limits
of stability.

From our calculations, a limiting asymmetry in the neutron-protron
concentration emerges beyond which a nucleus even within the gaseous
environment ceases to exist. A delicate balance between the Coulomb
force and the diluted surface tension with increasing asymmetry and
increasing density of the environment possibly plays a pivotal role
here. We also investigate the ambient conditions, namely the average
density, electron or proton fraction and the neutrino concentration,
under which nuclei of a particular species can be formed at different
temperatures in the stellar matter. We furthermore study the thermal
evolution of the nuclear clusters at different densities, which may
serve as a guide to understand the nucleation process of different
species of nuclear clusters from the nucleonic sea as the neutron star
cools down in the earlier stages of its formation.

We have left our calculations on a simple pedestal. That is, we have
not included the extension of the Thomas-Fermi framework, we have not
taken the shell-effects into account, possible plasma effects have
been ignored and, similarly, the presence of $\alpha$-particles at low
densities has not been taken into consideration. We have worked with
the SKM* interaction; at large asymmetry, its validity is not
unquestionable. Sophistications in the approach or the use of more
suitable interactions may change the results somewhat, but the
qualitative features that emerge from our calculations, we believe,
will remain mostly unchanged.

Two of us (X.V. and M.C.) acknowledge financial support from the
DGICYT (Spain) under grant PB98-1247 and from DGR (Catalonia) under
grant 1998SGR-00011.

%

\newpage
\centerline{\bf Figure Captions}
\begin{itemize}

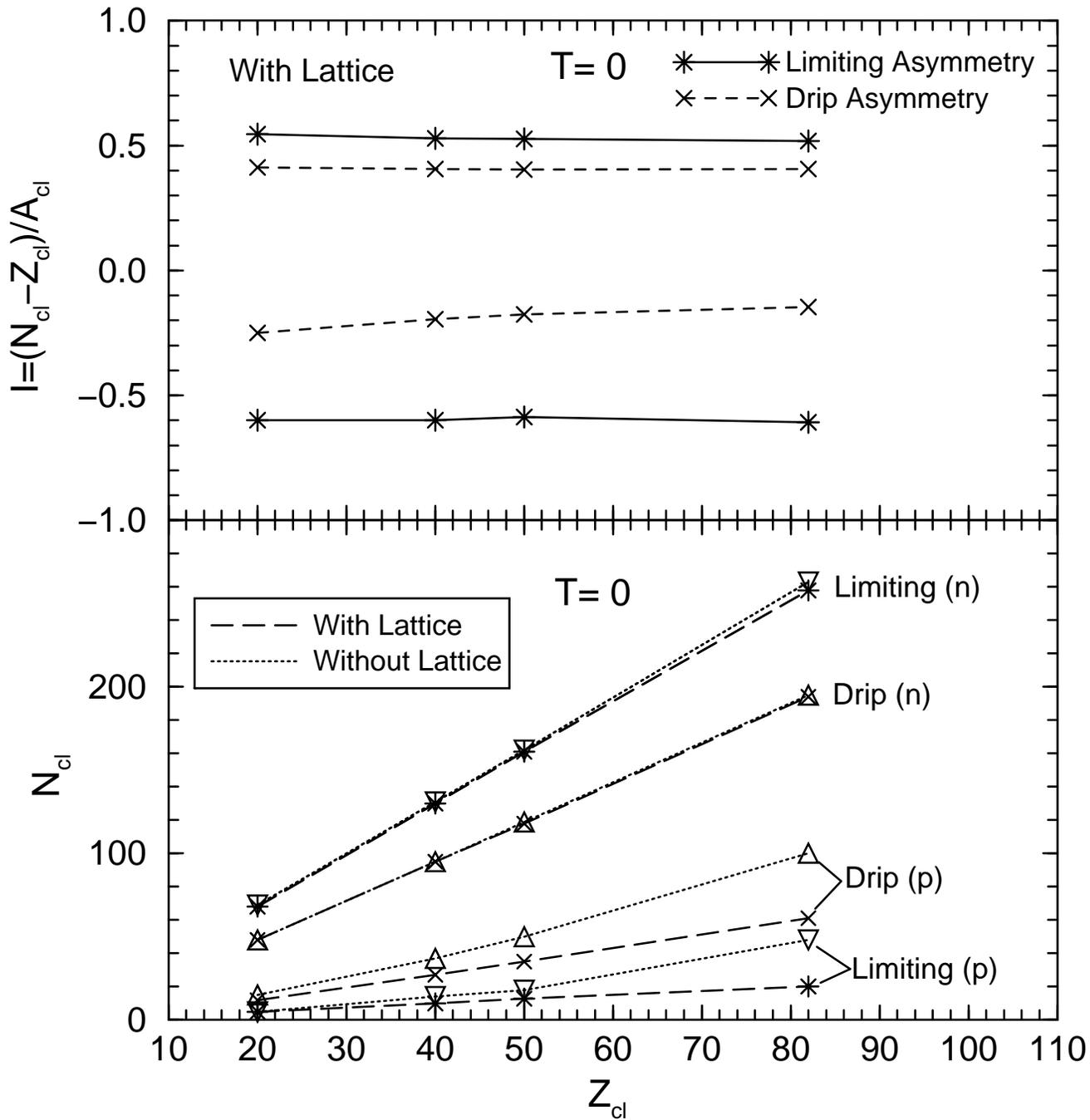
\item[Fig.\ 1] The upper panel shows the drip line asymmetry and the
limiting asymmetry as a function of the charge $Z_{cl}$ of the nuclear
cluster in a lattice of size $R_c=15$ fm at $T=0$. The lower panel
displays the same quantities in the $N_{cl}-Z_{cl}$ plane with and
without inclusion of lattice effects.

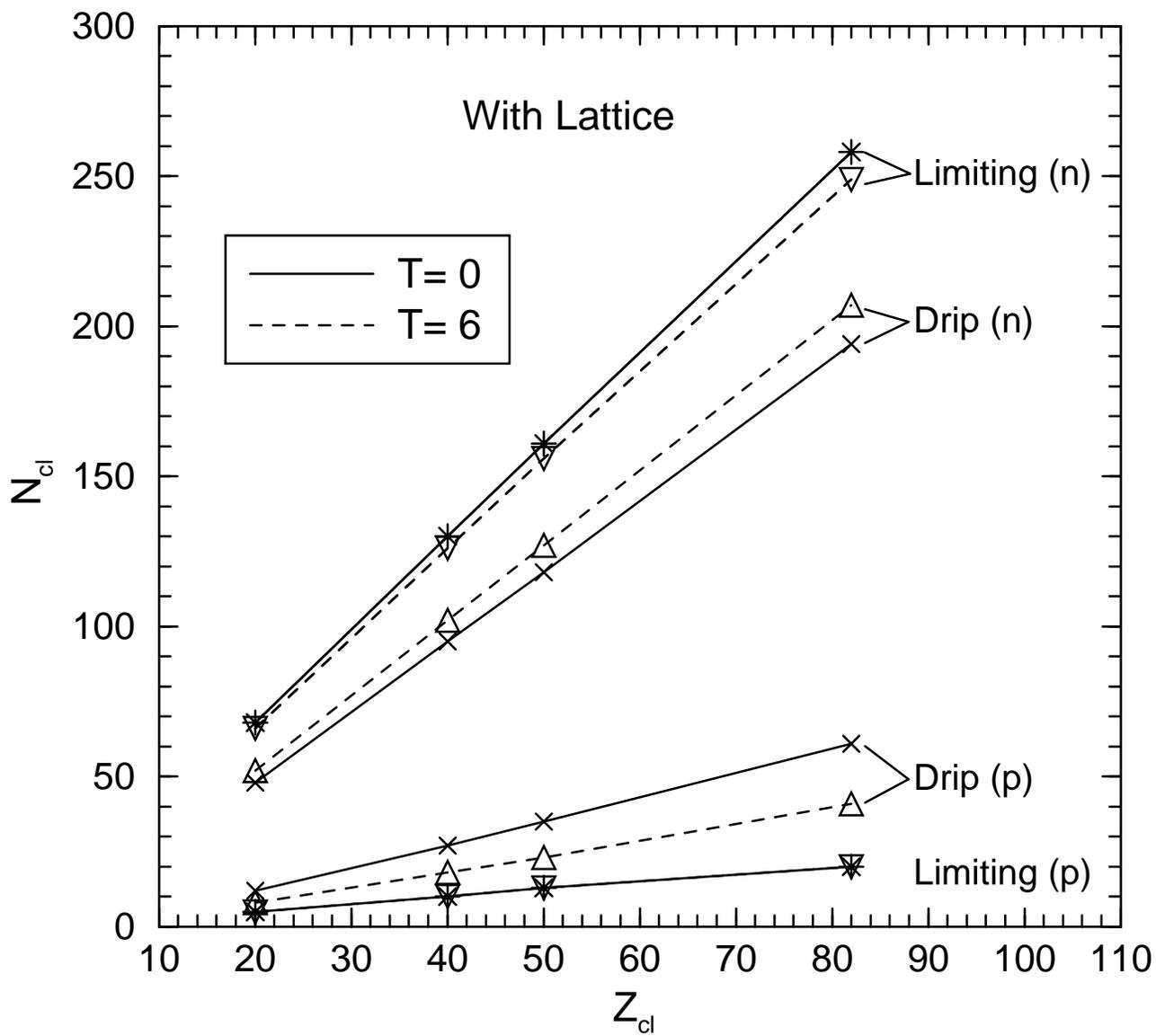
\item[Fig.\ 2] The drip and the limiting asymmetries in the
$N_{cl}-Z_{cl}$ plane at $T=0$ and 6 MeV in a lattice of $R_c=15$
fm. The limiting proton asymmetry coincides at both temperatures.

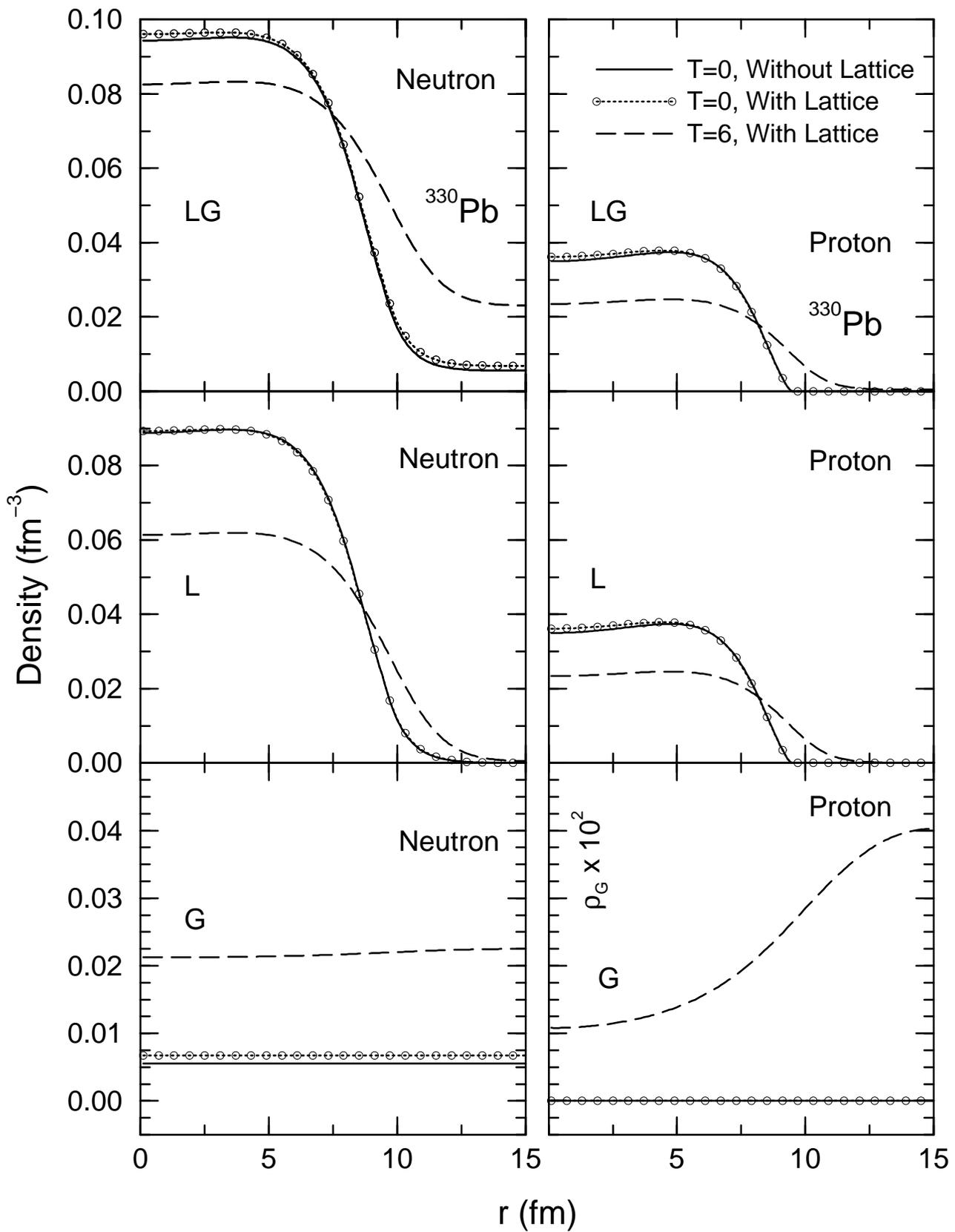
\item[Fig.\ 3] The neutron (left panel) and proton (right panel)
density profiles corresponding to the nuclear cluster $^{330}$Pb are
shown with inclusion of lattice effects at $T=0$ MeV (open circles
joined by dotted lines) and at $T=6$ MeV (dashed line). The full line
corresponds to calculations for the isolated nucleus (without lattice)
at $T=0$. The upper panels refer to the liquid-plus-gas density (LG)
and the central panels correspond to the liquid density (L). The lower
panels show the gas density (G).

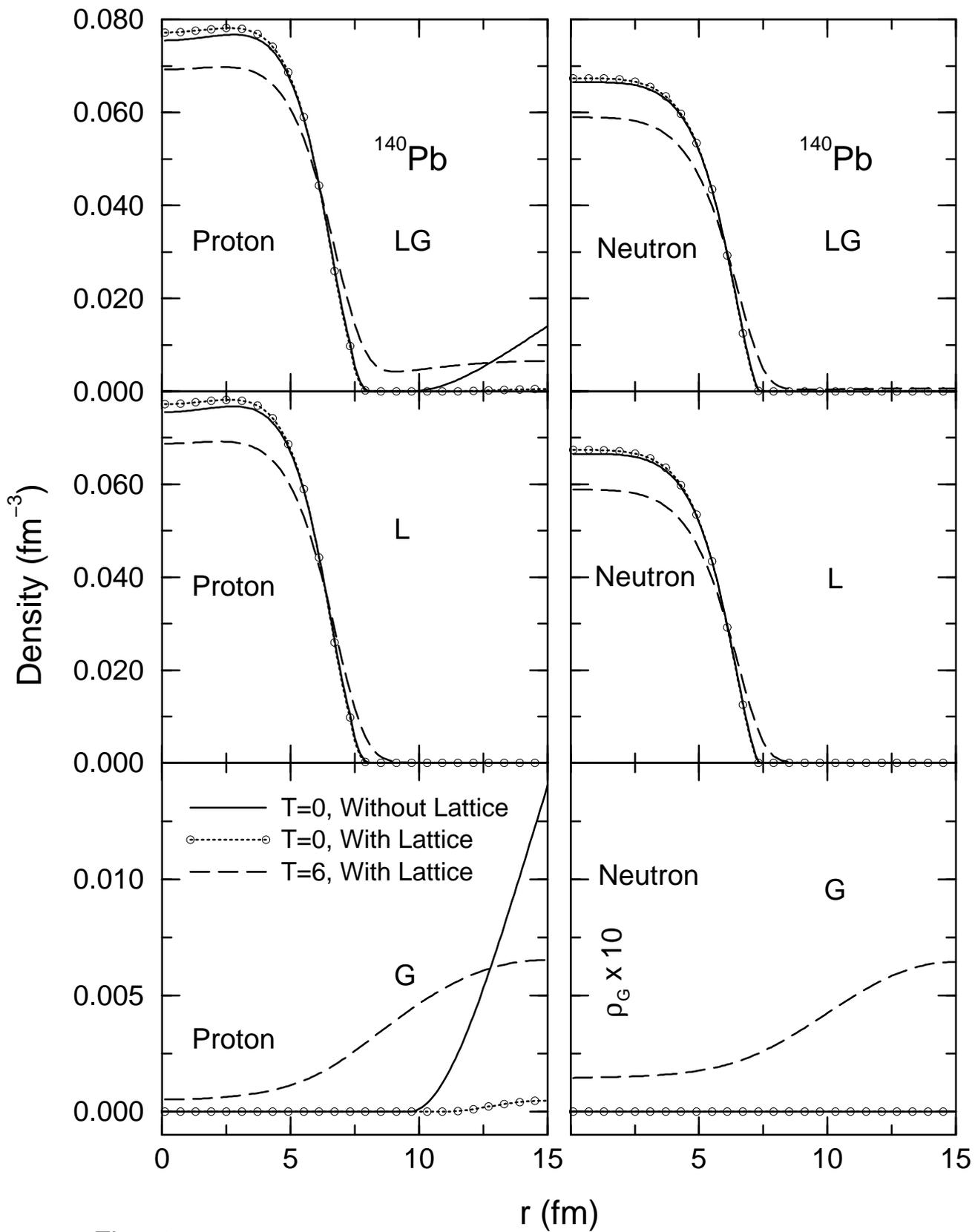
\item[Fig.\ 4] Same as in Fig.\ 3 for $^{140}$Pb. Here the left panel
corresponds to protons and the right panel refers to neutrons.

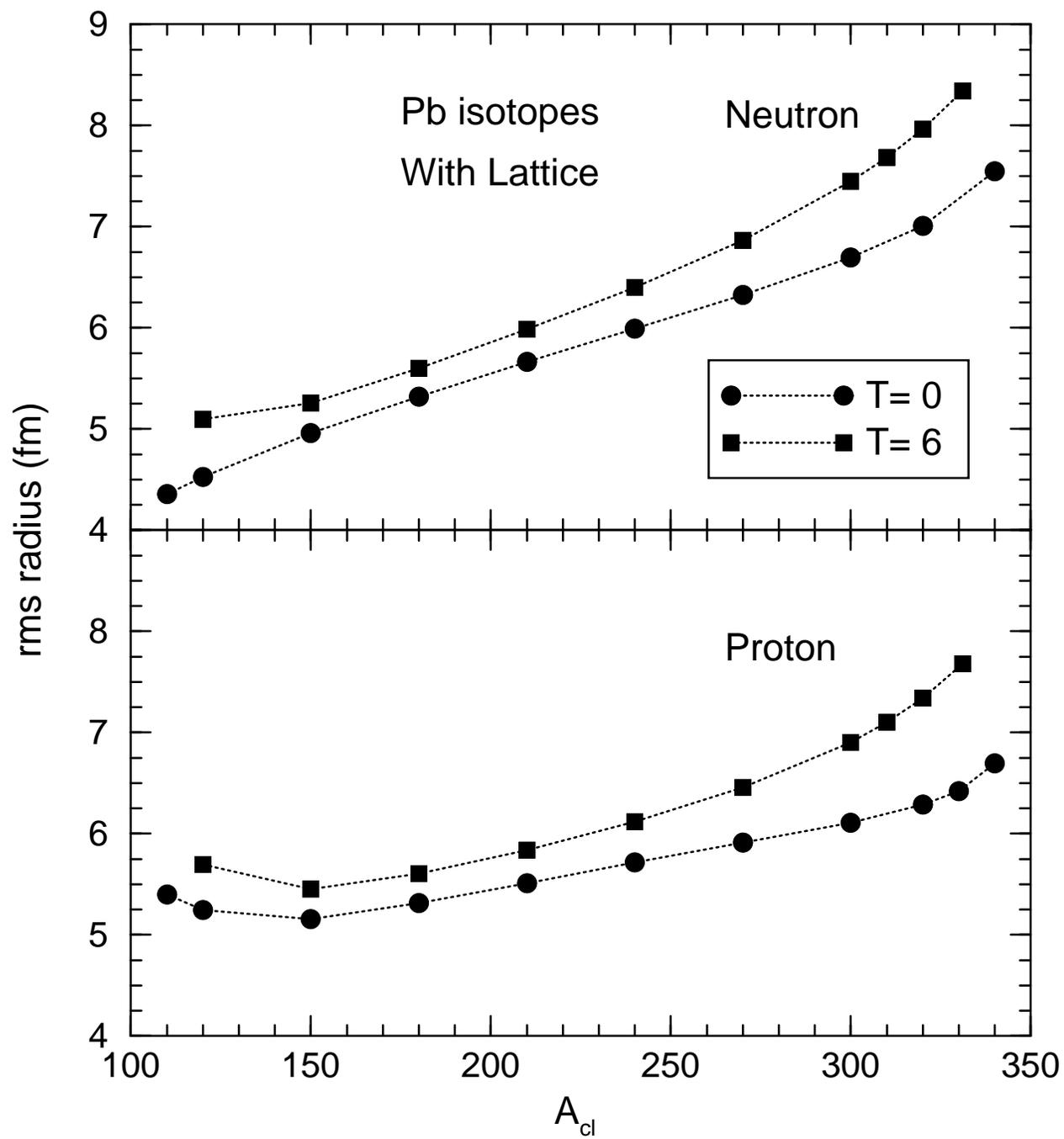
\item[Fig.\ 5] The neutron (top panel) and proton (bottom panel) rms
radii at $T=0$ and 6 MeV calculated in a lattice cell of size $R_c=15$
fm for Pb isotopes.

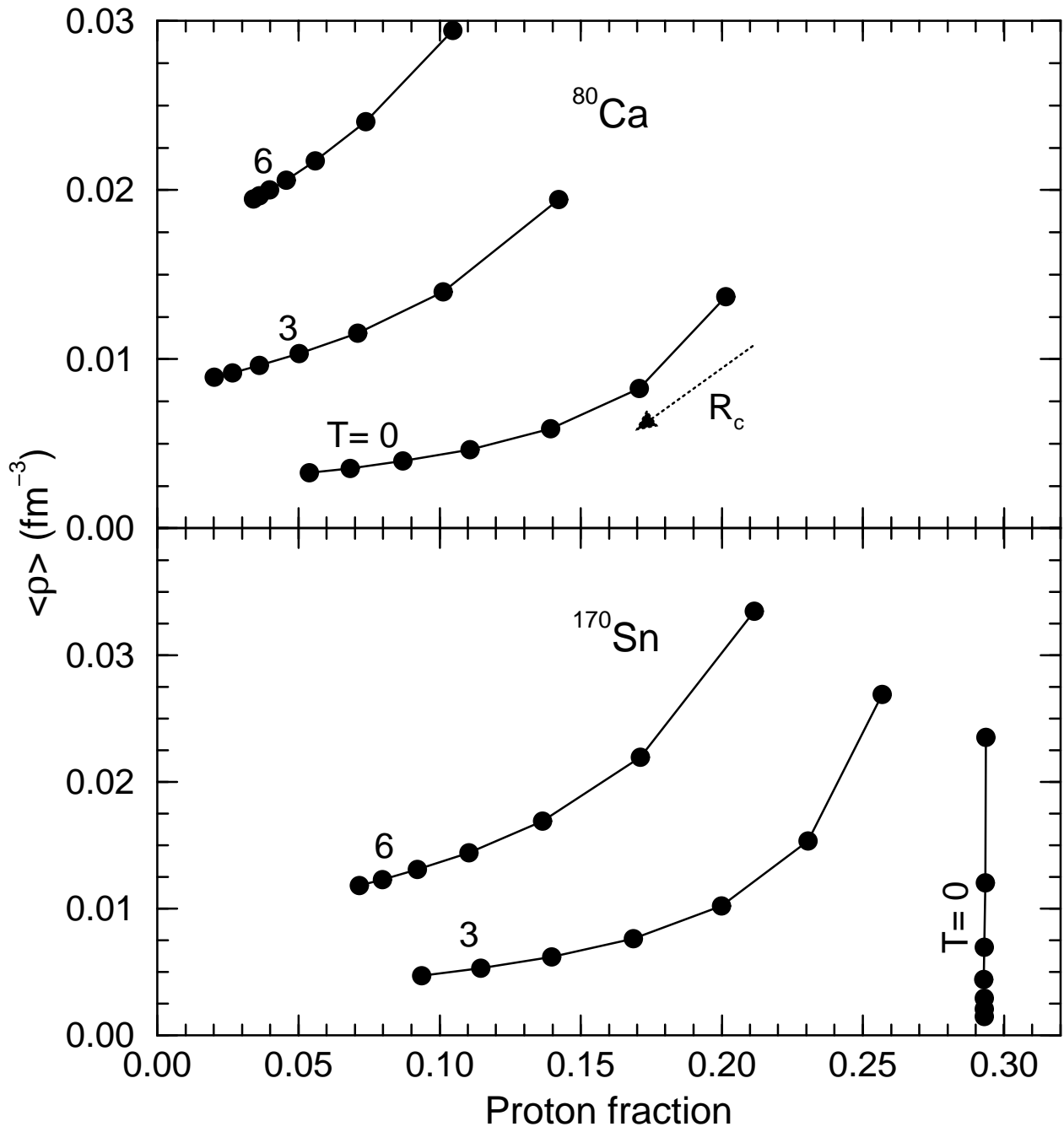
\item[Fig.\ 6] The average density and proton fraction of the nuclear
matter in the Wigner-Seitz cell calculated for the nuclear clusters
$^{80}$Ca and $^{170}$Sn. The size $R_c$ of the cell is varied from 12
fm to 30 fm at a step of 3 fm, as marked by the filled circles. The
calculations are done for temperatures $T=0$, 3 and 6 MeV\@. The full
lines are drawn to guide the eye.

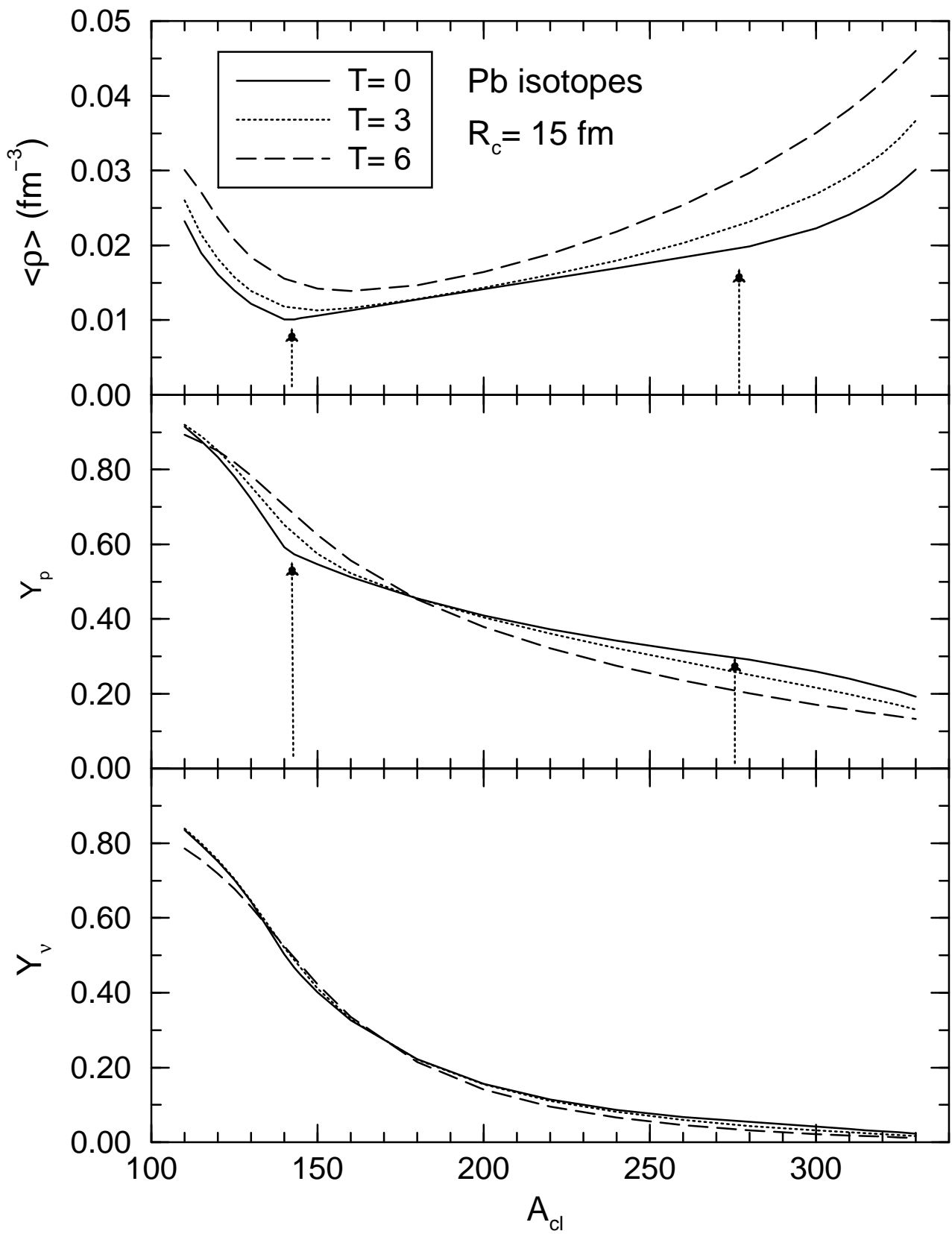
\item[Fig.\ 7] The variation of the average density
$\langle\rho\rangle$, the proton fraction $Y_p$ and the neutrino
fraction $Y_{\nu}$ as a function of the mass number $A_{cl}$ of the Pb
isotopes calculated in a Wigner-Seitz cell of size 15 fm. The
calculations are done at $T=0$, 3 and 6 MeV\@. The vertical arrows
indicate the proton and neutron drip points.

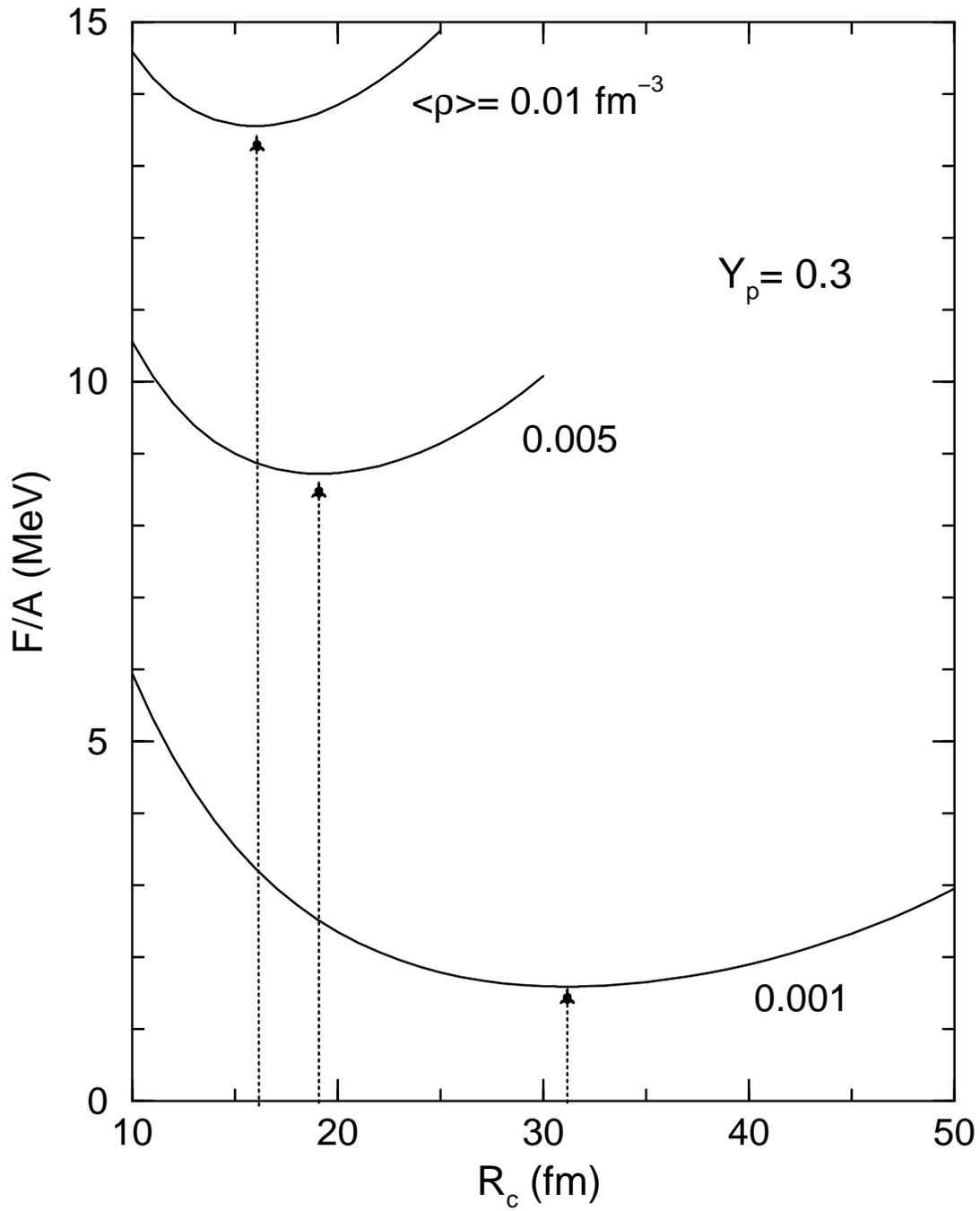
\item[Fig.\ 8] The variation of the free energy per baryon $F/A$ with
the cell size at average densities $\langle\rho\rangle=0.001$, 0.005
and 0.01 fm$^{-3}$, the proton fraction being fixed at $Y_p=0.3$. The
calculations correspond to $T=0$. The vertical arrows refer to the
minima in $F/A$.

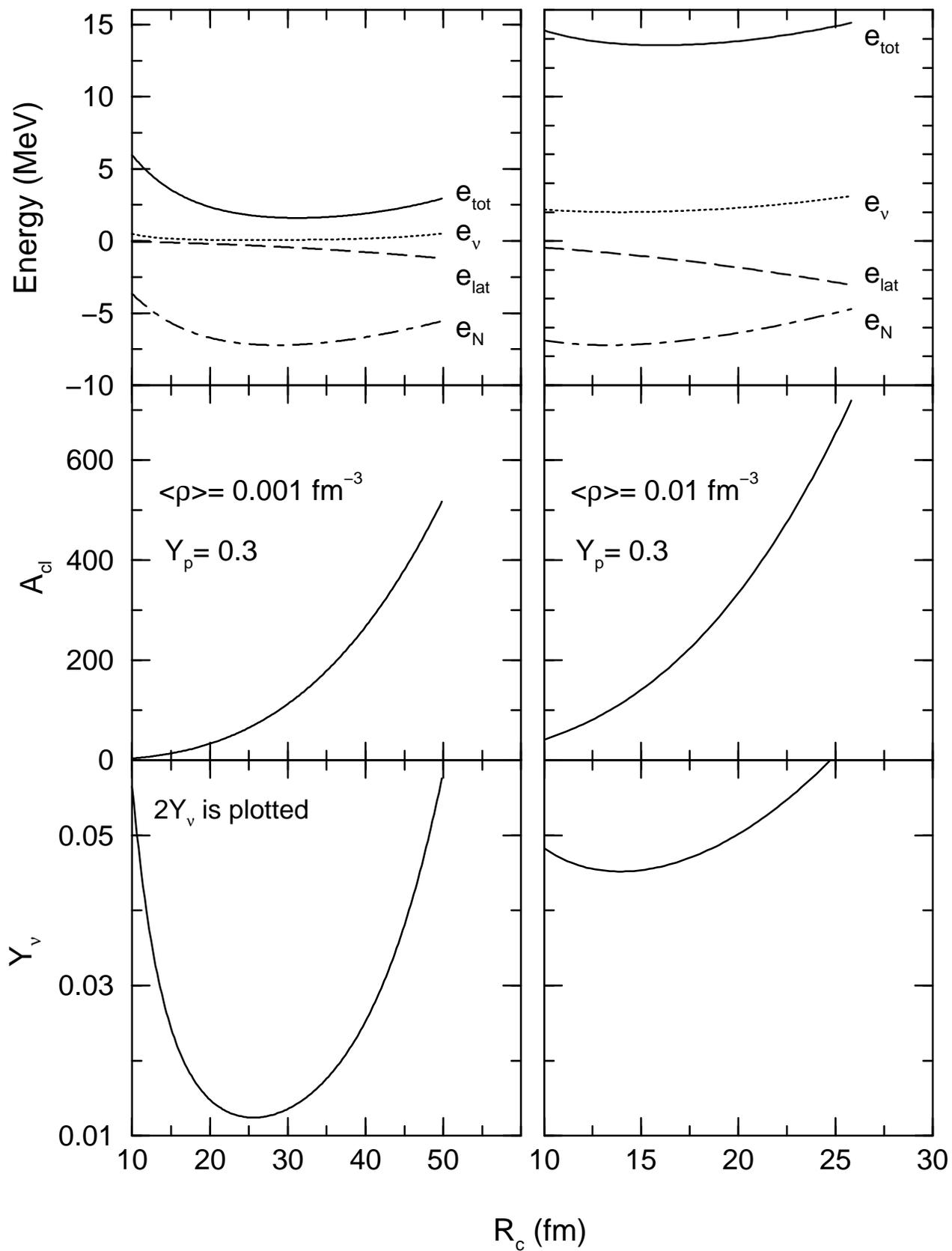
\item[Fig.\ 9] The different components of the free energy per nucleon
as a function of $R_c$ are shown in the upper panel for two different
values of $\langle\rho\rangle$, at $Y_p=0.3$ and $T=0$. The middle
panels show the growth of the nuclear cluster with cell size. The
bottom panels display the variation of $Y_\nu$ as a function of $R_c$.
The numbers on the left bottom panel are to be scaled down by a factor
of two.

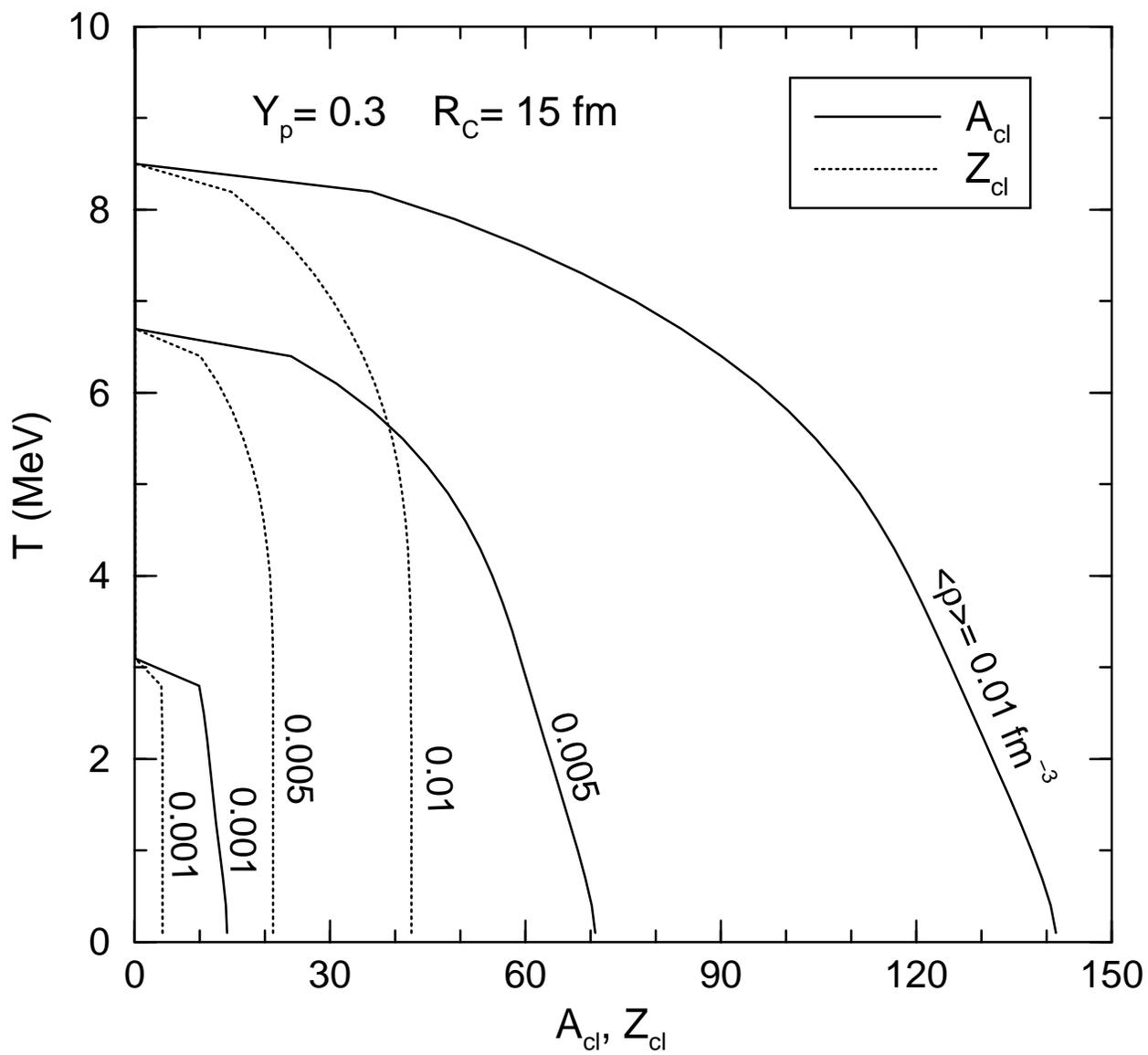
\item[Fig.\ 10] Thermal evolution of the cluster sizes for average
densities $\langle\rho\rangle=0.001$, 0.005 and 0.01 fm$^{-3}$. The
full lines correspond to the mass number $A_{cl}$ and the dotted
lines refer to the atomic number $Z_{cl}$. The calculations are
performed with a fixed proton fraction $Y_p=0.3$ and a cell size
$R_c=15$ fm.

\end{itemize}

\newpage
\begin{thebibliography}{99}
\bibitem{bay} G. Baym. H. A. Bethe, and C. J. Pethick, 
 Nucl. Phys. {\bf A175}, 225 (1971).
\bibitem{buc} J. R. Buchler and Z. Barkat, 
 Phys. Rev. Lett. {\bf 27}, 48 (1971).
\bibitem{pet} C. J. Pethick and D. G. Ravenhall, 
 Ann. Rev. Nucl. Part. Sci. {\bf 45}, 429 (1995).
\bibitem{hei} H. Heiselberg and M. Hjorth-Jensen,
 Phys. Rep. {\bf 328}, 237 (2000).
\bibitem{sha} S. L. Shapiro and S. A. Teukolsky, Black Holes, White
Dwarfs, and Neutron Stars, John Wiley \& Sons, New York (1983).
\bibitem{lan} W. D. Langer, L. C. Rosen, J. M. Cohen, and A. G. W. Cameron,
Astrophys. Space Sci. {\bf 5}, 529 (1969).
\bibitem{rav} D. G. Ravenhall , C. J. Pethick, and J. R. Wilson, Phys. Rev.
Lett. {\bf 50}, 2066 (1983).
\bibitem{oya} K. Oyamatsu, Nucl. Phys. {\bf A 561}, 431 (1993).
\bibitem{lor} C. P. Lorenz, D. G. Ravenhall, and C. J. Pethick, 
Phys. Rev. Lett. {\bf 70}, 379 (1993).
\bibitem{che} K. S. Cheng, C. C. Yao, and Z. G. Dai, Phys. Rev. {\bf C 55},
2092 (1997).
\bibitem{eng} L. Engvik, E. Osnes, M. Hjorth-Jensen, G. Bao, and E.
{\O}stgaard,  Astrophys. J. {\bf 469}, 794 (1996);
J. M. Lattimer and M. Prakash, Astrophys. J. {\bf 550} 426 (2001).
\bibitem{lat1} J. M. Lattimer and D. G. Ravenhall, 
               Astrophys. J. {\bf 223}, 314 (1978).
\bibitem{lam}  D. Q. Lamb, J. M. Lattimer, C. J. Pethick, 
 and D. G. Ravenhall, Nucl. Phys {\bf A 360}, 459 (1981).
\bibitem{mul} H. M\"uller and B. D. Serot, Phys. Rev. {\bf C 52},
2072 (1995).
\bibitem{lat} J. M. Lattimer, C. J. Pethick, D. G. Ravenhall, and D. Q.
Lamb, Nucl. Phys {\bf A 432}, 646 (1985).
\bibitem{mye} W. D. Myers, W. J. Swiatecki, and C. S. Wang, Nucl. Phys. {\bf
A 436}, 185 (1985).
\bibitem{cen} M. Centelles, M. Del Estal, and X. Vi\~nas, Nucl. Phys.
{\bf A 635}, 193 (1998).
\bibitem{bon} P. Bonche, S. Levit, and D Vautherin, Nucl. Phys. {\bf A 427},
278 (1984); Nucl. Phys. {\bf A 436}, 265 (1985).
\bibitem{sur} E. Suraud, Nucl. Phys. {\bf A 462}, 109 (1987).
\bibitem{jde} J. N. De, X. Vi\~nas, S. K. Patra, and M. Centelles,
Phys. Rev. {\bf C 64}, 057306 (2001).
\bibitem{mpi} M. Pi, X. Vi\~nas, M. Barranco, A. Perez-Canyellas,
and A. Polls, Astron. Astrophys. Suppl. Ser. {\bf 64}, 439 (1986).
\bibitem{bra} M. Brack, C. Guet, and H. B. H{\aa}kansson,
Phys. Rep. {\bf 123}, 275 (1985).
\bibitem{cox} J. P. Cox and R. T. Giuli, 
Principles of Stellar Structure, Gordon and Breach, New York (1968).
\bibitem{bes} J. Besprosvany and S. Levit,
Phys. Lett. {\bf B217}, 1 (1989).
\bibitem{jde1} J. N. De, D. Bandopadhyay, S. K. Samaddar, and N. Rudra, 
Nucl. Phys. {\bf A 534}, 294 (1991).
\bibitem{jde2} J. N. De, B. K. Agrawal, and S. K. Samaddar,
Phys. Rev.  {\bf C 59}, 1 (1999).
\end{thebibliography}
\end{document}